\documentclass{PoS}

\title{L\"uscher-Weisz algorithm for excited states of the QCD flux-tube}

\ShortTitle{L\"uscher-Weisz algorithm for excited states of the QCD flux-tube}

\author{\speaker{Bastian B. Brandt}, Pushan Majumdar \\%
Westf\"alische Wilhelms-Universit\"at M\"unster -- Institut f\"ur Theoretische Physik -- Wilhelm-Klemm-Str. 9, 48149 M\"unster\\%
E-mail: \email{bastianbrandt@uni-muenster.de}\\%
E-mail: \email{pushan@uni-muenster.de}}

\abstract{We present a version of the L\"uscher-Weisz multilevel algorithm ideally suited for studying excited 
states of the QCD flux tube. While the original version achieved error reduction only in the temporal direction, 
the new algorithm reduces fluctuations in the sources as well. We report on the implementation of this algorithm 
as well as improvement over the older method. We also present first results, where we see a good agreement 
with theoretical predictions from bosonic string models.}

\FullConference{The XXV International Symposium on Lattice Field Theory\\%
July 30 - August 4 2007\\%
Regensburg, Germany}%

\begin{document}

\section{Introduction}

Formation of a flux-tube between a quark and an antiquark in the QCD vacuum plays an important role in the 
description of the still unsolved phenomenon of quark confinement. At a large distance $ R $ between the 
quarks, it is believed that the QCD flux tube has quite a similar behavior to an oscillating bosonic string,
largely independent of the internal gluonic degrees of freedom which make up the flux tube. 

This led to the formulation of effective string theories for the flux tube \cite{God} and a lot of studies 
have dealt with the groundstate of the string \cite{pot}. The simplest model describing the excited states of the string is that of the free bosonic string.
 A more realistic model is provided by the Nambu string theory. 
An important development by \textit{Arvis} \cite{Arvis} was the writing down of a $q{\bar q}$ potential 
assuming the flux tube to be described by a Nambu string.  

Nevertheless the Nambu theory still has the conformal anomaly in anything other than 26 dimensions. This has led 
to several other proposals for effective string theories. \textit{Lüscher} and \textit{Weisz} \cite{Weisz2} imposed an open- closed duality on the string partition function and obtained the Nambu string spectrum up to order 
$ 1/R^{3} $.
Another effective string theory introduced by \textit{Polchinski} and \textit{Strominger} \cite{Polchinski} consists of the most general terms at every order in $ 1/R $, which does not introduce the conformal anomaly. In this theory too the string spectrum agrees with Arvis up to corrections of order $ 1/R^{5} $ \cite{Dass}. For a recent review see \cite{Kuti}.

In this article we look at the excitation spectrum of the flux tube formed between a static quark and an 
antiquark in 3 dimensional SU(2) lattice gauge theory and compare our results with the predictions for open 
bosonic string spectra. For a similar study in the closed string sector see \cite{Teper}.

\section{Sources and excited states}

In 2+1 dimensions, the energy states of the oscillating string can be classified by parity and charge-conjugation properties,
$(C,P)$, and one can identify the ground states in these channels with the lowest four energy states of the string:
\begin{equation}
\label{eq1}
E_{0} \leftrightarrow (+,+) \quad E_{1} \leftrightarrow (+,-) \quad E_{2} \leftrightarrow (-,-) \quad E_{3} \leftrightarrow (-,+)
\end{equation}
Our goal is now to measure these energy states on the lattice.

The groundstate of the potential can be measured well with Polyakov loop correlators, which have the spectral 
representation
$
\left\langle P ( R , T ) \: P ( 0 , T ) \right\rangle = \sum_{n=0}^{\infty} b_{n} \: \exp \left[ - E_{n} (R) \: T \right] .
$
In this case the coefficients $ b_{n} $ are integers and $ E_{n} (R) $ are the energy states at quark separation 
$ R $. Effective string prediction for ground state $ V(R) = E_{0} (R) $ is
\begin{equation}
V(R) = - \lim_{T\to\infty} \frac{1}{T} \: \ln \left\langle P(R,T) \: P(0,T) \right\rangle = V_{0} + \sigma \: R - \frac{\pi}{24} \: ( d - 2 ) \: \frac{1}{R} + \dots
\end{equation}
where $V_0$ is an unphysical constant and $\sigma$ is the string tension.
The $ 1/R $ term is the well known L\"uscher-term \cite{Luescher}, reproduced by all effective string theories. 
Different models however give different predictions for the excitation spectrum and therefore one way to distinguish between different string-models is to measure the energy-differences between the excited states.

\begin{figure}
\centering
\begin{minipage}[c]{0.40\textwidth}
\includegraphics[width=.8\textwidth]{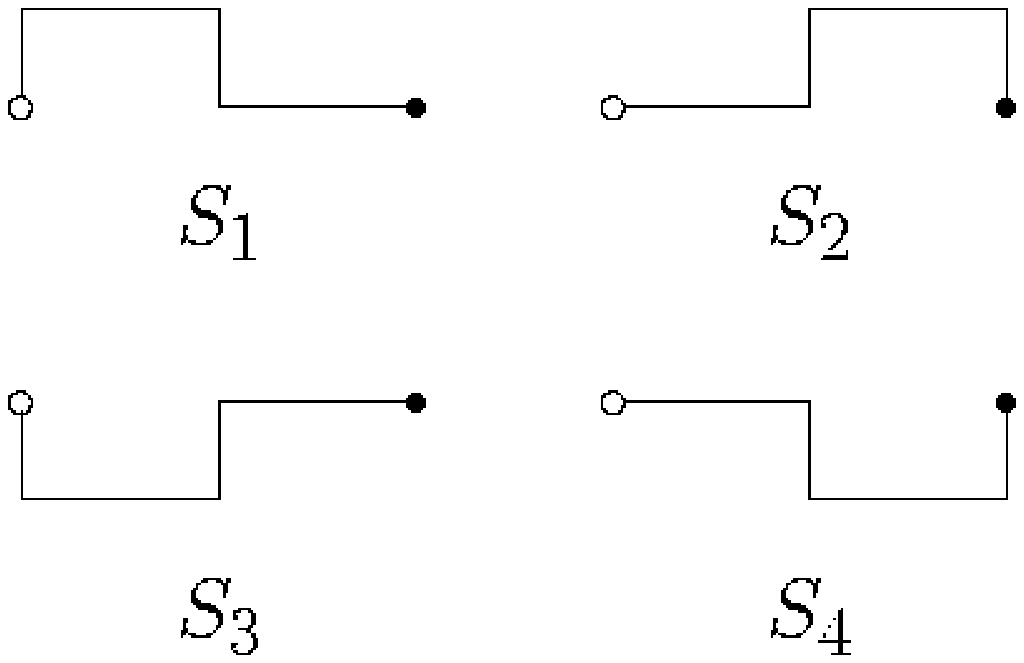}
\end{minipage}
\begin{minipage}[c]{0.40\textwidth}
\includegraphics[width=.8\textwidth]{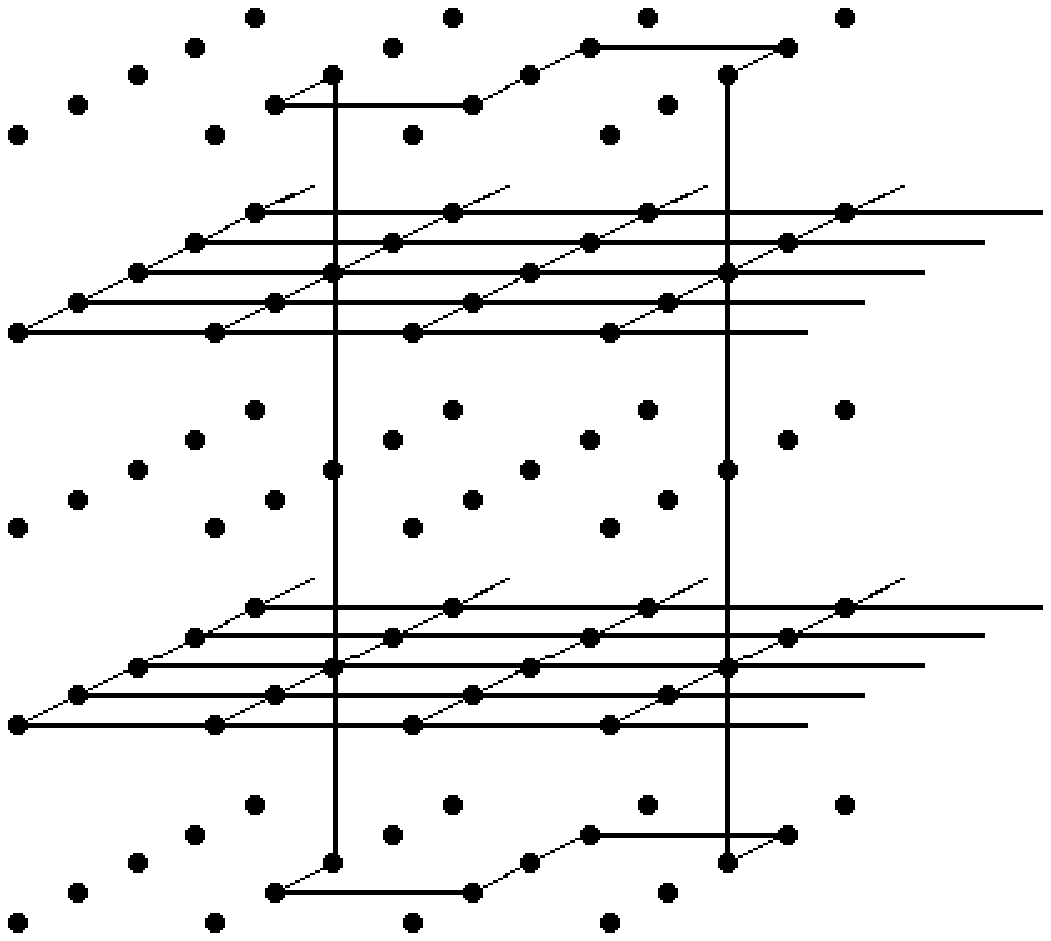}
\end{minipage}
\caption{\textbf{Left:} Spatial lines that correspond to the sources on the lattice. \textbf{Right:} A Wilson loop with sources at the ends, that lie in the middle of the time slices. The slices with the solid lines are the time slices with fixed lines during the sublattice updates.}
\label{pic1}
\end{figure}

The Polyakov loop correlators used to calculate the groundstate of the potential are bad estimators for the excited states.
A better way of measuring the excited states is provided by Wilson loops. However normal Wilson loops, with straight 
spatial lines at the ends, will again project strongly on the ground state, but weakly on the excited states. 
In order to get a preferential coupling to the excited states, we use a set of wavefunctions at the ends of the 
loops, called sources. These sources correspond to spatial lines on the lattice, that replace the straight 
spatial lines of the Wilson loops. The set of sources we use is shown in figure (\ref{pic1}).

If one creates superpositions of these sources with well defined parity and charge conjugation, these `channels' 
couple directly to the excited states according to equation \ref{eq1}. The superpositions are shown in table (\ref{tab:par}).

\begin{table}
\begin{center}
\small
\begin{tabular}{|c|r|c|c|r|cc|cr|}
\cline{1-3}\cline{5-9}
&&&& $T$ & $\:$ $R=4-9$ $\:$ & & $R=10-12$ & \\
channel & superposition~~~~ & state &&& sources & time tr. & sources & time tr.~~\\
\cline{1-3}\cline{5-9}
$ (+,+) $ & $ S_{1} + S_{2} + S_{3} + S_{4}~ $ & $ E_{0} $ & \hspace*{6mm}& 4 & 12000 & 1000 & 24000 & 1000~ \\
$ (+,-) $ & $ S_{1} + S_{2} - S_{3} - S_{4}~ $ & $ E_{1} $ & \hspace*{6mm}& 6 & 12000 & 1500 & 24000 & 2000~ \\
$ (-,-) $ & $ S_{1} - S_{2} - S_{3} + S_{4}~ $ & $ E_{2} $ & \hspace*{6mm}& 8 & 12000 & 2000 & 24000 & 6000~ \\
$ ~(-,+)~ $ & $ ~- S_{1} + S_{2} - S_{3} + S_{4}~ $ & $ E_{3} $ & \hspace*{6mm}& $12 $ & 12000 & 2500 & 24000 & 12000~\\
\cline{1-3}\cline{5-9}
\end{tabular}
\normalsize
\end{center}
\caption{{\bf Left :} Combination of the sources into $(C,P)$ eigenstates. {\bf Right :} Run parameters for the 
multilevel algorithm we used.}\label{tab:par}
\end{table}

There have been other attempts to measure the excited states. \textit{Juge}, \textit{Kuti} and \textit{Morningstar}
\cite{Juge} used asymmetric lattices with small physical temporal extent and a large number of basis states to 
measure the excited states in different field theories. In that respect our approach is complementary to theirs, 
as we use relatively few basis states, but use much larger physical temporal extents to exponentially suppress the
contamination due to excited states.

We can now calculate the expectation values of Wilson loops $ W^{K} (R,T) $ with channel $ K $ at the ends, which
has the representation:
\begin{equation}
\label{eq2}
W^{K}(R,T) = \alpha^{K} \: e^{ - E^{K} (R) \: T } \: \left( 1 + \beta^{K} \: e^{ - \Delta E^{K} (R) \: T } + \dots \right)
\end{equation}
Thus we are able to measure the different energy states with the formula:
\begin{equation}
E^{K} (R) = - \frac{1}{T_{2}-T_{1}} \: \ln \left[ \frac{W^{K}(R,T_{2})}{W^{K}(R,T_{1})} \right] + corrections
\end{equation}
The ``$corrections$" in this expression are due to the higher states in the channel and, as can be seen from equation (\ref{eq2}) are exponentially damped with $ T_{1} $ and $ T_{2} $. This is why one would like to go to Wilson loops with large time extends.

The problem with large Wilson loops is of course the very small signal to noise ratio.

Even for the numerical cheap $SU(2)$ LGT in d=2+1, conventional methods do not work. One way of reducing the error is provided by the L\"uscher-Weisz algorithm \cite{Weisz1}, which leads to an exponential error reduction for the time transporters of the Wilson loops. Putting the sources on fixed time-slices of the lattice and using the sublattice updates to reduce the fluctuations of the time transporters produces good results and is practicable for loops with time extends up to about a Fermi, see \cite{Pushan2}.

However this is not the best way to use the algorithm because the fluctuations of the sources are not reduced with this method.

\section{The new method}

To achieve further error reduction we now move the sources from the fixed lines to the middle of the time slices. Such a Wilson loop with sources at the ends is also shown in figure (\ref{pic1}).
\\
Main advantages of such a procedure are:
\vspace*{-2mm}

\begin{itemize}
 \item The fluctuations of the sources are reduced by the sublattice updates as well. \vspace*{-2mm}
 \item One can use multihit on single links of the sources that leads to a further error reduction. \vspace*{-2mm}
\end{itemize}

It is also beneficial to use different numbers of sublattice updates for the time slices that contain the sources and the time slices that contain only time-transporters. In this way it is possible to choose parameters for the algorithm to optimize the noise to speed ratio for the single parts of the Wilson loops.

Several tests show, that it is good to use more sublattice updates for the sources than for the time transporters for excited states. If one chooses the right parameters for the sources and the time slices, one is able to achieve an error reduction of $ \mathcal{O} ( 10 ) $ for the same computation time.

A related algorithm was used by \textit{Kratochvila} and \textit{de Forcrand} to look at string breaking with Wilson loops \cite{forcrand}.

\section{First results}

In our first run we worked on a $ 24^{3} $-lattice for $ R=4-9 $ and on a $ 48^{3} $-lattice for $ R = 10-12 $, 
with $ \beta = 5 $ ($r_{0}=3.9536(3)$), made 2000 total measurements using the scheme of sublattice updates 
shown in table (\ref{tab:par}). Compared to \cite{Pushan2} we were able to increase the time extent of the loops from $ T=2,4,6,8 $ to $ T=4,6,8,12$.

\begin{figure}
\centering
\includegraphics[angle=-90, width=11.8cm]{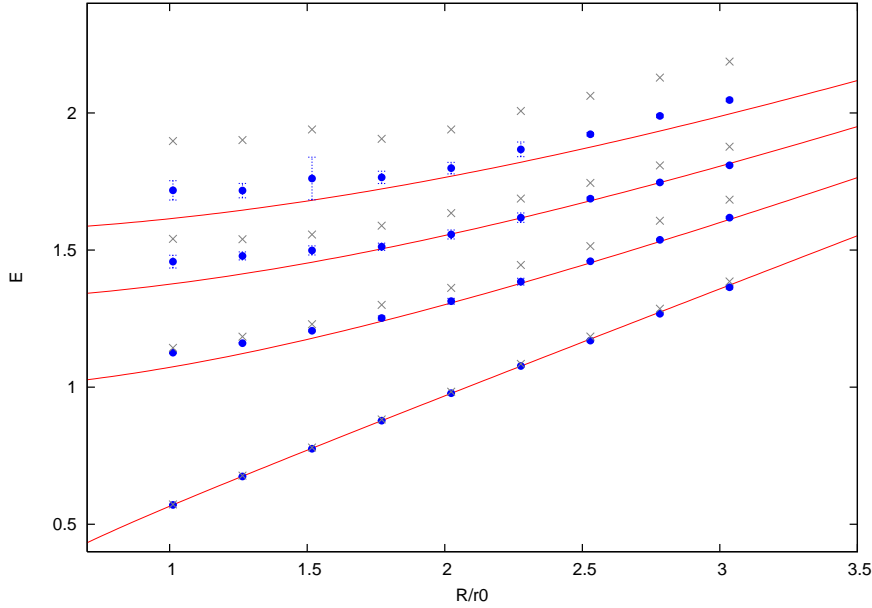}
\caption{Measured energy values and predictions of the Arvis potential: The grey crosses are the naive 
energy values and the blue points are the corrected ones. The red lines are predictions of the Arvis
potential.}
\label{fig-r1}
\end{figure}

The naive energies are calculated with the formula:
\begin{equation}
E_{\alpha}(R) = - \frac{1}{T_{2}-T_{1}} \: \ln \left[ \frac{ W_{\alpha} (R,T_{2}) } { W_{\alpha} (R,T_{1}) } \right]
\qquad {\rm with}\quad T_1 \,\, \textnormal{and} \,\, T_2 = 4 \,\, \textnormal{and} \,\, 8
\end{equation}
To get rid of the contamination due to the higher states in the channels, we use a fit to the form
\begin{equation}
- \frac{1}{T_{2}-T_{1}} \: \ln \left[ \frac { W_{\alpha} ( R , T_{2} ) } { W_{\alpha} ( R , T_{1} ) } \right] = \bar{E}_{\alpha} ( R ) + \frac{1}{T_{2}-T_{1}} \: \left[ a \: e^{ - b \: T_{1} } \: \left( 1 - e^{ - b \: ( T_{2} - T_{1} ) } \right) \right] ,
\end{equation}
with the fit parameters $ \bar{E}_{\alpha} , b , c $ to calculate the corrected energy $ \bar{E}_{\alpha} $. For 
the corrected energies $ \bar{E}_{2} $ and $ \bar{E}_{3} $ we were only able to use Wilson loops with the time 
extends $ T=4,6,8 $. For the first two states we used Wilson loops with all time extents. The values we obtained 
are given in table(\ref{tab:en}).

\begin{table}
\begin{center}
{ \footnotesize
\begin{tabular}{|l|cc|cc|cc|cc|}
\hline
$R$ & $E_{0}$ & $\bar{E}_{0}$ & $E_{1}$ & $\bar{E}_{1}$ & $E_{2}$ & $\bar{E}_{2}$ & $E_{3}$ & $\bar{E}_{3}$ \\
\hline
4 & 0.5725(3) & 0.571(1) & 1.143(2) & 1.126(3) & 1.541(9) & 1.46(3) & 1.90(8) & 1.72(4) \\
5 & 0.6776(4) & 0.674(2) & 1.184(2) & 1.161(3) & 1.540(8) & 1.48(2) & 1.90(7) & 1.72(3) \\
6 & 0.7801(6) & 0.775(2) & 1.230(2) & 1.206(4) & 1.557(8) & 1.50(2) & 1.94(9) & 1.76(8) \\
7 & 0.8826(9) & 0.8781(6) &1.300(3) & 1.252(8) & 1.589(9) & 1.51(2) & 1.91(6) & 1.77(3) \\
8 & 0.983(2) & 0.9779(8) & 1.362(3) & 1.31(2) & 1.635(10) & 1.56(2) & 1.94(7) & 1.80(3) \\
9 & 1.085(2) & 1.0772(10) & 1.446(4) & 1.38(2) & 1.69(1) & 1.62(2) & 2.01(8) & 1.87(3) \\
10 & 1.1847(4) & 1.170(5) & 1.5146(8) & 1.459(3) & 1.745(3) & 1.688(6) & 2.06(2) & 1.923(7) \\
11 & 1.2862(4) & 1.268(4) & 1.6070(9) & 1.537(4) & 1.809(3) & 1.747(5) & 2.13(2) & 1.989(7) \\
12 & 1.3858(5) & 1.364(6) & 1.6841(10) & 1.618(4) & 1.877(4) & 1.809(5) & 2.19(2) & 2.047(6) \\
\hline
\end{tabular}
}
\end{center}
\vspace*{-1mm}
\caption{Naive and corrected energies for the four lowest states in the different $(C,P)$ channels.}\label{tab:en}
\vspace*{-1mm}
\end{table}

\begin{figure}
\centering 
\includegraphics[angle=-90, width=11.8cm]{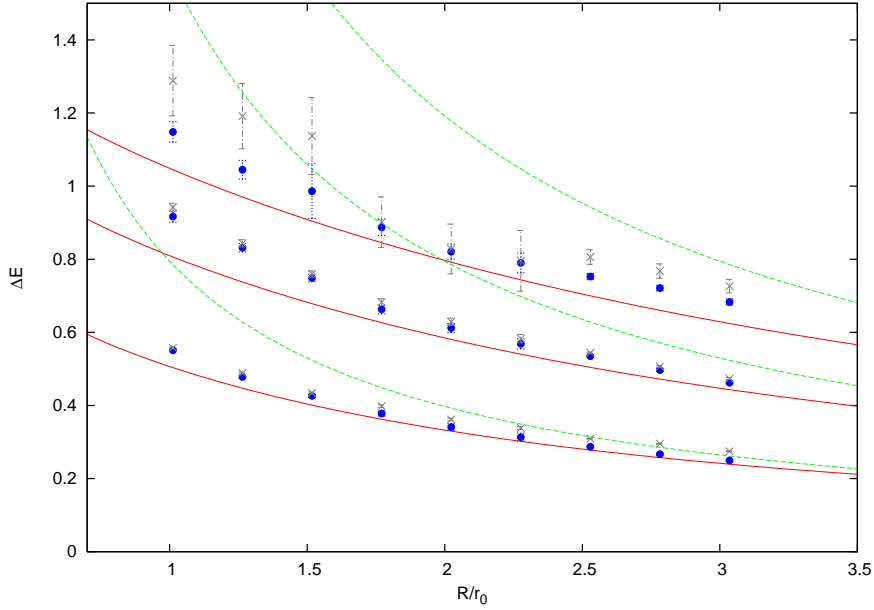}
\caption{The energy differences $ \Delta \bar{E}_{10} $, $ \Delta \bar{E}_{20} $ and $ \Delta \bar{E}_{30} $ and the prediction of the Arvis potential: The grey crosses are again the naive differences and the blue points the measured values. The red lines are the predictions of the Arvis potential and the green lines are the predictions of the free bosonic string.}
\label{fig-r2}
\end{figure}

Even more interesting than the total energy values are the energy differences. Again to take the corrections into account, we use a three-parameter fit
\begin{equation}
- \frac{1}{T_{2}-T_{1}} \: \ln \left[ \frac { W_{n} ( R , T_{2} ) \: W_{0} ( R , T_{1} ) } { W_{n} ( R , T_{1} ) \: W_{0} ( R , T_{2} ) } \right] = \Delta \bar{E}_{n0} + \frac{1}{T_{2}-T_{1}} \: b \: e^{ - c \: T_{1} } \: \left( 1 - e^{ - c \: ( T_{2} - T_{1} ) } \right) ,
\end{equation}
with the parameters $ \Delta \bar{E}_{n0} , b , c $. The corresponding values are shown in table (\ref{tab:ediff}).
\begin{table}
\begin{center}
{\footnotesize
\begin{tabular}{|l|ccccccccc|}
\hline
$R$ & 4 & 5 & 6 & 7 & 8 & 9 & 10 & 11 & 12 \\
\hline
$ \Delta \bar{E}_{10} $ & 0.551(4) & 0.478(5) & 0.426(4) & 0.378(7) & 0.341(10) & 0.31(2) & 0.287(3) & 0.267(3) & 0.250(3) \\
$ \Delta \bar{E}_{20} $ & 0.92(2) & 0.83(2) & 0.75(1) & 0.66(2) & 0.61(2) & 0.57(2) & 0.294(2) & 0.273(3) & 0.256(3) \\
$ \Delta \bar{E}_{30} $ & 1.15(3) & 1.04(3) & 0.99(8) & 0.89(3) & 0.82(2) & 0.79(3) & 0.753(9) & 0.721(8) & 0.683(9) \\
\hline
\end{tabular}
}
\end{center}
\caption{Corrected energy differences of the excited states with the ground state.}\label{tab:ediff}
\end{table}

Finally in figure (\ref{fig-r1}), we plot the total energies and in figure (\ref{fig-r2}) the energy difference 
against the predictions of the Arvis potential \cite{Arvis}, which gives the energy states
\begin{equation}
E_{n}(R) = \sigma \: R \: \sqrt{ 1 + \frac{2\pi}{\sigma\:R^{2}} \: \left( n - \frac{1}{24} \: ( d - 2 ) \right) } .
\end{equation}

\section{Conclusions}

We have discussed an algorithm that allows us to use the advantage of the L\"uscher-Weisz algorithm for the time transporters as well as for the sources and we presented our first results using the algorithm. We were able to go to much bigger Wilson loops than was possible previously.
We see from figure (\ref{fig-r2}) that at the same $ q \bar{q} $ separation $R$ the energy difference $ \Delta \bar{E}_{10} $ is much closer to the predictions from \textit{Arvis} than the energy differences $ \Delta \bar{E}_{20} $ and $ \Delta \bar{E}_{30} $, which agrees with the expactations in \cite{Weisz2}. Also the magnitude of the corrections seems to be larger for the higher states.
Unfortunately the error reduction obtained with the naive states is not sufficient to get a signal for the third
excited state beyond $T=8$. At the moment it is not clear whether this is due to lack of statistics or poor choice
of parameters in the multilevel algorithm. Investigation into this issue is ongoing.

\acknowledgments
All results were obtained employing the condor sytem \footnote{\textbf{Copyright:} 1990-2007 Condor-Team,
Computer Sciences Department, University of Wisconsin -- Madison, WI.} \cite{condor} to make efficient use of the 
computing resources
of the Westfälische Wilhelms-Universität Münster. We are thankful to the university for this facility.

\end{document}